\documentclass[11pt,a4paper]{article}
\usepackage{hyperref}
\usepackage[super,sort,compress]{cite}

\usepackage{graphicx}
\usepackage{dcolumn}
\usepackage{bm}
\usepackage[usenames,dvipsnames]{xcolor} 
\usepackage[utf8]{inputenc}
\usepackage{latexsym}
\usepackage{amsfonts}
\usepackage{amssymb}
\usepackage{amsthm}
\expandafter\let\csname equation*\endcsname\relax 
\expandafter\let\csname endequation*\endcsname\relax
\usepackage{amsmath}
\usepackage{psfrag}
\usepackage{rotating}
\usepackage{pgf,pgfsys,pgffor} %
\usepackage{pgfplots}
\usepackage{pgfplotstable}
\usepackage{tikz}
\usetikzlibrary{intersections,arrows,decorations.pathmorphing,backgrounds,fit,mindmap,calc,through,scopes,fadings,positioning,automata,calendar,shapes,er,matrix,folding,patterns,petri,plothandlers,plotmarks,shadows,topaths,through,trees,pgfplots.units,decorations, snakes}  
\usepgfplotslibrary{fillbetween}

\usepackage{caption}
\usepackage{subcaption}

\usepackage{lineno,hyperref}
\modulolinenumbers[5]

\newcommand{\trace}[1]{\textrm{Tr} \left[ {#1} \right]}
\newcommand{\bra}[1]{\langle {#1} \vert}
\newcommand{\ket}[1]{\vert {#1} \rangle}
\newcommand{\pure}[1]{\vert {#1} \rangle \langle {#1} \vert}

\newcommand{\ee}{e} 
\newcommand{\dd}{\textrm{d}}	
\newcommand{\ii}{{i\mkern1mu}}				

\newcommand{\beq}{\begin{equation}}
\newcommand{\eeq}{\end{equation}}

\usepackage{authblk}

\begin{document}

\title{Quantum Stochastic Walk Models \\ for Quantum State Discrimination}

\author{Nicola Dalla Pozza\footnote{Electronic address: nicola.dallapozza@unifi.it (Corresponding author)}, Filippo Caruso\footnote{Electronic address: filippo.caruso@unifi.it}}
\affil{Dipartimento di Fisica e Astronomia, \\
Universit\`a degli Studi di Firenze, \\
I-50019 Sesto Fiorentino, Italy}

\date{}

\maketitle

\begin{abstract}
Quantum Stochastic Walks (QSW) allow for a generalization of both quantum and classical random walks by describing the dynamic evolution of an open quantum system on a network, with nodes corresponding to quantum states of a fixed basis. We consider the problem of quantum state discrimination on such a system, and we solve it by optimizing the network topology weights. Finally, we test it on different quantum network topologies and compare it with optimal theoretical bounds. 
\end{abstract}





\section{Introduction}

Quantum Stochastic Walks (QSW) have been proposed as a framework to incorporate decoherence effects into quantum walks (QW), which, on the contrary, allow only for a purely unitary evolution of the state \cite{Aharonov1993, Kempe2003, VenegasAndraca2012}. Unitary dynamics has been sufficient to prove the computational universality \cite{Childs2009, Childs2013} and the advantages provided by QW for quantum computation and algorithm design \cite{Childs2009, Childs2013, Farhi1998, Chids2003}, 
but it is worthwhile the possibility to include effects of decoherence. 
Indeed, the beneficial impact of decoherence has already been proved in a variety of systems \cite{Plenio2008, MendozaArenas2013, ContrerasPulido2014}, in particular in light-harvesting complexes \cite{Caruso2009,Chin2010, Caruso2010}. As a consequence, shortly after their formalization QSW have been investigated in relation to propagation speed \cite{Domino2017, Domino2018}, learning speed-up \cite{Schuld2014}, steady state convergence \cite{SanchezBurillo2012, Liu2017} 
and enhancement of excitation transport \cite{Mohseni2008, Caruso2014, Viciani2015, Caruso2016, Park2015}.

Effectively, this framework allows to interpolate between quantum walks and classical random walks \cite{Whitfield2010}. The evolution of QSW is defined by a Gorini– Kossakowski–Sudarshan–Lindblad 
master equation  \cite{Kossakowski1972, Lindblad1976, Gorini1976}, written as 
\beq
\frac{\dd \rho}{\dd t}= -(1-p)\ii \left[ H,\rho \right] + p \sum_{k} \left( L_{k} \rho L_{k}^{\dagger} - \frac{1}{2} \left \{ L_{k}^{\dagger}L_{k}, \rho \right \} \right).
\label{eq:Lindblad}
\eeq
In \eqref{eq:Lindblad}, the Hamiltonian $H$ accounts for the coherent evolution, the set of Lindblad operators $L_k$ accounts for the irreversible evolution, while the smoothing parameter $p$ defines a linear combination of the two. For $p=0$ we obtain a quantum walk, for $p=1$ we obtain a classical random walk (CRW).

We consider the problem of discriminating a set of known quantum states $\{\rho^{(n)}\}_n$ prepared with a priori probabilities $\{p_n\}_n$. By optimizing the set of  Positive Operator-Valued Measure $\{\Pi_n\}_n$ we aim at estimating the prepared quantum state with the highest probability of correct detection $P_c = \sum_{n} p_n \trace{\Pi_n \rho^{(n)}}$. In its general formulation the problem requires numerical methods \cite{Eldar2003}, but close analytical solutions are  available in the case of symmetric states \cite{Eldar2004, Nakahira2013, DallaPozza2015}. For instance, in the binary discrimination of pure states the optimal $P_c^{(opt)}$ is known as Helstrom bound and it evaluates $P_c^{(opt)}=(1+\sqrt{1-4p_1p_2 \trace{\rho^{(1)}\rho^{(2)}}})/2$. For a review of the results of quantum state discrimination in quantum information theory we refer to \cite{Helstrom1976, Holevo1972, Yuen1975, Chefles2000, Bergou2007, Bergou2010, Barnett2009}, and to \cite{Fanizza2019} for a recent connection with machine learning. 
In here we consider the problem in a quantum system represented by a network, and we test different models 
optimizing the discrimination on multiple topologies.

\section{\label{sec:networks}Quantum Stochastic Walks on networks}

We consider a quantum system which is represented by a network $\mathcal{G}~=~(\mathcal{N},\mathcal{E})$, where the nodes $\mathcal{N}_i \in \mathcal{N}$ corresponds to the quantum states of a fixed basis and the links  $\mathcal{N}_i \to \mathcal{N}_j \in \mathcal{E}$ depend on the hopping rates between the nodes. As in classical graph theory, the adjacency matrix $A$ indicates whether a link is present ($A_{j,i}=1$), or not ($A_{j,i}=0$).
In \emph{weighted} graphs, $A_{i,j}$ is a weight associated to the link. 
A symmetric adjacency matrix $A_{i,j} = A_{j,i}$ refers to an \emph{undirected} graph, otherwise the graph is said to be \emph{directed}.

In this paper we will focus on the continuous version of random walks. CRWs are defined with a transition-probability matrix $T$ which represents the possible transitions of a walker from a node onto the connected neighbours. From the adjacency matrix it is possible to define $T=AD^{-1}$, where $D$ is the (diagonal) degree matrix, with $D_i = \sum_j A_{j,i}$ representing the number of nodes connected to $i$. The probability distribution of the node occupation, written as a column vector ${\vec q}(t)$, is evaluated for a continuous time random walk as $\frac{\dd {\vec q}}{\dd t} = (T-I) {\vec q}$.

Quantum walks are simply defined by posing $H=A$, and defining the evolution as 
$ \frac{\dd \rho}{\dd t} = - \ii \left[ H,\rho \right]$. The population on the nodes is obtained applying a projection 
on the basis of the Hilbert space associated with the nodes.

In the case of a QSW, the dynamics is defined from Eq. \eqref{eq:Lindblad} with $H$ and $L_k = L_{i,j} =  \sqrt{T_{i,j}} \ket{i}\bra{j}$, with $0\leq T_{i,j} \leq 1$, $\sum_i T_{i,j}=1$. While this is sufficient to define a proper Lindblad equation, and it also gives the correct limiting case for $p \to 0$ and $p \to 1$, there are different ways to define $H$ and $T$ from $A$. For instance, in both \cite{Caruso2014, Whitfield2010} we have $H = A,\ T=AD^{-1}$, with the difference that in \cite{Caruso2014} the weights can only be unitary or null. Here we want to compare the performance of these models with the case where $H$ and $T$ are related with $A$ by $A_{i,j} = 0 \Longrightarrow H_{i,j} = 0,\  T_{i,j} = 0$, with $H$ being a real symmetric matrix with null diagonal and $T$ verifying $0\leq T_{i,j} \leq 1$, $\sum_i T_{i,j}=1$. Alternatively, we can think that $A_{i,j}$ is either 1 or 0, and it acts as a marker on the links that we want to optimize (1) or switch off (0).
This scheme relaxes the constraints on $H_{i,j},\  T_{i,j}$, allowing for additional degrees of freedom to exploit in the discrimination.

\section{Network model}

To define the quantum system, we consider a network organized in layers, which mimics the structure of neural networks \cite{Bishop, Goodfellow, HastieTibshiraniFriedman}, with $M$ input nodes, $N$ intermediate ancillary nodes and $O$ output nodes in the model $M-N-O$ (see Fig.~\ref{fig:MNO}).
The quantum system is initialized only on the input nodes with $\rho(0) \in \{\rho^{(m)}\}_{m=1}^\mathcal{M}$.
Thus, input nodes define a sub-space of size $M$ used to access the network, with $M$ in general not related to $\mathcal{M}$.
The quantum system then evolves according to \eqref{eq:Lindblad} through the intermediate nodes
and into the output nodes. 
Such nodes are \emph{sink} nodes where the population gets trapped, that is, the network realizes an irreversible one-way transfer of population from a \emph{sinker} node $s_n$ to the $n$-th sink via the operator $L_n=\ket{n}\bra{s_n}$. We add the sum total
\beq
2 \Gamma_s \sum_{n=1}^\mathcal{M} \ket{n}\bra{s_n}\rho \ket{s_n} \bra{n} - \frac{1}{2} \left \{ \pure{s_n}, \rho \right \}
\label{LindbladSink}
\eeq
on the right--hand side of Eq. \eqref{eq:Lindblad}, setting $\Gamma_s=1$ in our simulations. We consider a sinker node for each sink, but more sinkers connected to the same sink could be introduced. However, the number of sinks $O$ must be equal or greater than $\mathcal{M}$ to have a sink for each hypothesis on the prepared states. In fact, at the end of the time evolution a measurement is performed by projecting on the nodes basis, and if the outcome $n$ corresponding to the $n$-th sink is obtained we estimate that the quantum state $\rho^{(n)}$ has been prepared. The network will be optimized such that these estimations works as best as possible, and if an outcome corresponding to a node that is not a sink occurs, we consider it inconclusive.

\tikzstyle{inputNode}=[circle,thick,draw=blue,minimum size=8mm, node distance=15mm]
\tikzstyle{hiddenNode}=[circle,thick,draw=orange,minimum size=8mm, node distance=15mm]
\tikzstyle{outputNode}=[circle,thick,draw=purple,minimum size=8mm, node distance=15mm]
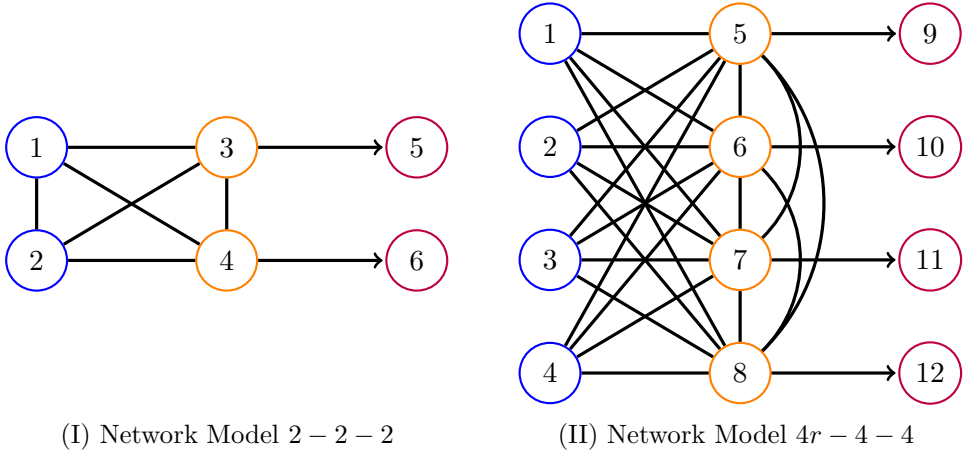
\begin{figure}[hbt]
\subcaptionbox{Network Model $2-2-2$ \label{fig:2-2-2}}{%
	\centering
	\begin{tikzpicture}[scale=.5, line width=1.2pt] 
	
		\node[inputNode] (A) {1};
		\node[inputNode, below of=A] (B) {2}
			edge (A);
		\node[hiddenNode, right of=A, xshift=1cm] (D) {3}
			edge (A)	edge (B);			
		\node[hiddenNode, right of=B, xshift=1cm] (E) {4}
			edge (A)	edge (B) edge (D);
		\node[outputNode, right of=D, xshift=1cm] (H) {5}
			edge [pre] (D);
		\node[outputNode, right of=E, xshift=1cm] (I) {6}
			edge [pre] (E);						
	\end{tikzpicture} 
	\vspace{1.5cm}
}%
\hfill
\subcaptionbox{Network Model $4r-4-4$ \label{fig:4r-4-4}}{%
	\centering
	\begin{tikzpicture}[scale=.5, line width=1.2pt]  
		\node[inputNode] (A) {1};
		\node[inputNode, below of=A] (B) {2};
		\node[inputNode, below of=B] (C) {3};
		\node[inputNode, below of=C] (D) {4};
		\node[hiddenNode, right of=A, xshift=1cm] (E) {5}
			edge (A)	edge (B) edge (C) edge (D);
		\node[hiddenNode, right of=B, xshift=1cm] (F) {6}
			edge (A)	edge (B) edge (C) edge (D) edge (E);
		\node[hiddenNode, right of=C, xshift=1cm] (G) {7}
			edge (A)	edge (B) edge (C) edge (D) edge (F) edge [bend right=45] (E);
		\node[hiddenNode, right of=D, xshift=1cm] (H) {8}
			edge (A)	edge (B) edge (C) edge (D) edge (G) edge [bend right=45] (E) edge [bend right=45] (F);
		\node[outputNode, right of=E, xshift=1cm] (I) {9}
			edge [pre] (E);
		\node[outputNode, right of=F, xshift=1cm] (L) {10}
			edge [pre] (F);
		\node[outputNode, right of=G, xshift=1cm] (M) {11}
			edge [pre] (G);	
		\node[outputNode, right of=H, xshift=1cm] (N) {12}
			edge [pre] (H);
	\end{tikzpicture} 
}%
\caption{Examples of $M-N-O$ network models. In each subfigure, the left layer (blue circles) collects the input nodes, in the right layer (purple circles) the output nodes and in between (orange circles) the intermediate nodes. Directed links specify irreversible processes towards the sinks, plain lines indicate non-zero entries in $A_{i,j}$. By default, all the nodes in a layer are connected with all the other nodes in the same layer and with the next one, except for the sink nodes that are connected only through their sinker node. When considering a \emph{reduced} connectivity in the graph, we indicate a $r$ near the number of nodes, as in (II). 
}
\label{fig:MNO}
\end{figure}

\section{Results}
We consider the two network models represented in Fig. \ref{fig:MNO}
and setup the optimization of the parameters in $H,\ T$ using four QSW schemes: (a) $H_{i,j} \geq 0,\ T=H D^{-1}$ as in \cite{Whitfield2010}, (b) $H_{i,j} \in \{0,1\},\ T=H D^{-1}$  as in \cite{Caruso2014}, (c) $H_{i,j} =-\max \{T_{i,j}, T_{j,i}\}, \ H_{i,i} = - \sum_{j\neq i} H_{j,i}$ and $T$ un--normalized as in \cite{Falloon2017} and (d) $H,\ T$ independently optimized, with the optimization variables indicated by $A$.

For the model $2-2-2$, we discriminate between the pure state $\rho^{(1)}$ and the mixed state $\rho^{(2)}$, written in the input sub-space basis $\ket{1}={1 \choose 0}, \ket{2}={0 \choose 1}$ as
\beq
\rho^{(1)}=\begin{pmatrix}
	\frac{2+\sqrt{2}}{4} & \frac{1 + \ii}{4} \\
	\frac{1 - \ii}{4} & \frac{2-\sqrt{2}}{4}
\end{pmatrix},
\rho^{(2)} = \begin{pmatrix}
	0.68 & -0.13 - 0.13 \ii \\
	-0.13 + 0.13 \ii & 0.32
\end{pmatrix}, p_1=p_2.
\notag
\eeq
In the case of the model $4r-4-4$, we discriminate between four equally probable quantum states $\rho^{(m)} = (1-\alpha)\frac{I}{4}+\alpha \pure{\varphi_m}$, defined as a linear combination of the completely mixed state and a pure state $\ket{\varphi_m}= \sum_{k=1}^4 \frac{\ee^{-\ii \frac{2 \pi m k}{4}}}{2} \ket{k}$, with $m=1, \ldots 4$ being the $m$-th state in the mutually unbiased basis of the input nodes $\ket{k}$. For $\alpha=1$ we would have the discrimination of the pure states $\ket{k}$, which would have a theoretical bound $P_c^{(opt)}=1$ since the states are orthogonal. We take $\alpha=0.7$ to simulate a noisy preparation of $\ket{\varphi_m}$.

\begin{figure}[hbt]
\subcaptionbox{Network Model $2-2-2$ \label{Pc:2-2-2}}{%
	\centering
	\includegraphics[width=0.48\columnwidth]{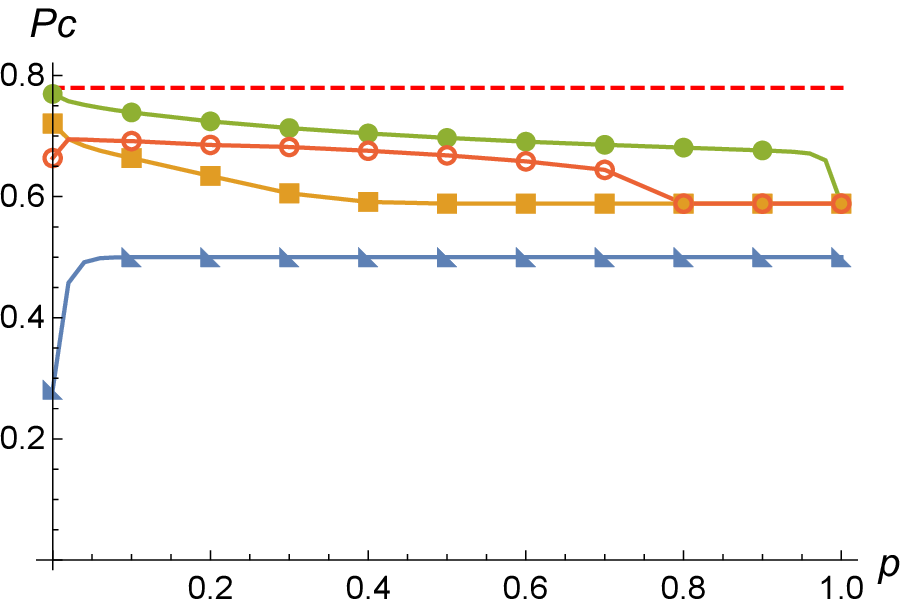}
}%
\hfill
\subcaptionbox{Network Model $4r-4-4$ \label{Pc:4r-4-4}}{%
	\centering
	\includegraphics[width=0.48\columnwidth]{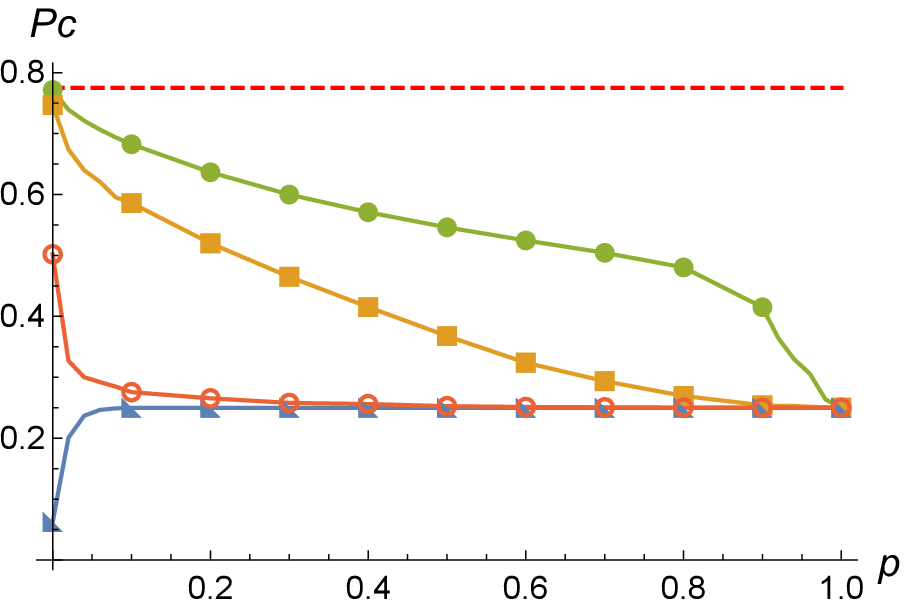}
}%
\caption{
Asymptotic probability of correct detection ($\tau=10^2s$) for scheme (a) in yellow (squared markers), (b) in blue (triangular markers), (c) in red (circular markers) and (d) in green (full disks). In red dashed line the value of $P_c^{(opt)}$.
}
\label{fig:performance}
\end{figure}

The optimal probability of correct decision can be evaluated 
using semi-definite programming \cite{Eldar2003}, and evaluates to $P_c^{(opt)} = 0.7795$ in the binary case, and to $P_c^{(opt)} = 0.7750$ in the $\mathcal{M}$-ary discrimination. 
We run the optimization for $p \in [0,1]$ and for multiple values of the total evolution time $\tau$. As expected, for increasing values of $\tau$ the performances also increase since more population can be transferred from the input nodes to the sink nodes. However, the performances saturate, and we plot the asymptotic behaviour as a function of $p$ in Fig. \ref{fig:performance}. 
For lower $\tau$ the trend in $p$ is similar.
\section{Discussion and conclusions}
We have considered the problem of discriminating a set of quantum states prepared in a quantum system whose dynamics is described by QSW on a network. We have investigated four different schemes to define the GKSL master equation from the network, and optimized the coefficients of the hamiltonian $H$ and the Lindblad operator coefficients $T$ as a function of $p$ and the total evolution time $\tau$. We have reported the asymptotic probability of correct detection, where a clear gap can be seen amongst the four schemes. The setup (d) with $H,\ T$ independently optimized gives the best performance on the whole range of $p$ due to the increased number of degrees of freedom but at the expenses of a higher computational cost for the optimization. Further investigations should consider how the performance scales in the number of nodes and layers to identify the best network topology, which may also depend on the quantum states to discriminate.
Finally, our results could be tested through already experimentally available benchmark platforms such as photonics-based architectures \cite{Caruso2016} and cold atoms in optical lattices \cite{DErrico2013}.
 


%
\section*{Acknowledgments}
This work was financially supported from Fondazione CR Firenze through the project Q-BIOSCAN and Quantum-AI, PATHOS EU H2020 FET-OPEN grant no. 828946, and UNIFI grant Q-CODYCES.
\bibliography{sqwDiscrimination_noUrl}

\end{document}